# Poverty, Unemployment and Displacement in Ukraine: three months into the war[1]

Maksym Obrizan,
Associate Professor, Kyiv School of Economics (KSE)
http://orcid.org/0000-0002-0924-0671

**Abstract**
This paper identifies the causal effects of full-scale kremlin aggression on socio-economic outcomes in Ukraine three months into the full-scale war. First, forced migration after February 24th, 2022 is associated with an elevated risk of becoming unemployed by 7.5% points. Second, difference-in-difference regressions show that in regions with fighting on the ground females without a higher education face a 9.6-9.9% points higher risk of not having enough money for food. Finally, in the regions subject to ground attack females with and without a higher education, as well as males without a higher education are more likely to become unemployed by 6.1-6.9%, 4.2-4.7% and 6.5-6.6% points correspondingly. This persistent gender gap in poverty and unemployment, when even higher education is not protective for females, calls for policy action. While more accurate results may obtain with more comprehensive surveys, this paper provides a remarkably robust initial estimate of the war's effects on poverty, unemployment and internal migration.

## 1. Introduction

A completely unbelievable and unprovoked full-scale kremlin aggression in Ukraine, which has had its own issues with economic development and corruption but has preserved democracy and press freedom, came as a cold shower surprise to the whole civilized world. Despite several episodes of heightened social confrontation, Ukraine has peacefully survived the collapse of the Soviet Union unlike the former Soviet republics of Moldova, Armenia, Azerbaijan, Georgia, Tajikistan, Uzbekistan, Kyrgyzstan, and russia which experienced military conflicts of different scale in the late 1980s and early 1990s (Obrizan and Iavorskyi 2022).

Although the conflict with russia started in 2014 after the annexation of the Ukrainian Autonomous Republic of Crimea and sponsoring of terrorist groups in Donbas, it has not then directly affected other regions of Ukraine except for large flows of internal migration and killed and wounded Ukrainian defenders from those regions. The situation tragically changed in the early morning of February 24th, 2022, with massive air strikes on critical infrastructure all over Ukraine (including the capital of Kyiv) and a ground attack from belarus, russia and Crimea along multiple dimensions.

Thousands of victims of bloody massacres by russian troops in Bucha[2], Irpin'[3], Mariupol'[4], Izium, Olenivka[5] and many other places call for justice for kremlin war criminals in Hague.[6] Likewise, russian federation will pay billions of USD in reparations for destroyed infrastructure in Ukraine.[7] But in addition to direct physical damage to people's lives and their health and destroyed infrastructure, the war led to a massive forced displacement, poverty and unemployment in Ukraine, not to mention the far-reaching consequences of increasing energy and food prices putting at risk the most vulnerable populations worldwide.

The economic consequences of the full-scale russian aggression are unprecedented. World Bank predicts Ukraine's GDP to decrease by 35% in 2022 under the current status quo without additional economic shocks.[8] Although it is challenging to estimate the impact of war on poverty, the baseline projection assumes Ukraine's poverty rate based on the USD 5.50 per day threshold will increase from 1.8% in 2021 to 19.8% in 2022 (World Bank 2022) but earlier estimates from the United Nations Development Programme (UNDP) suggest that up to 90% of Ukrainian citizens could be facing poverty and extreme economic vulnerability if the war deepens.[9] The International Labour Organization (ILO) estimates that 4.8 million jobs have been lost with respect to the pre-conflict situation, which is equivalent to 30% of pre-war employment in Ukraine.[10]

---

[1] I would like to thank the team of Kyiv International Institute of Sociology (KIIS) and, especially, Yulia Sakhno and Victoria Zakhozha for providing access to the data. All remaining errors are mine. Email for correspondence: mobrizan@kse.org.ua

[2] https://www.theguardian.com/world/2022/oct/20/ukraine-true-detectives-investigators-closing-in-on-russian-war-crimes
[3] https://uacrisis.org/en/irpin-massacre
[4] https://www.atlanticcouncil.org/blogs/ukrainealert/putins-mariupol-massacre-is-one-the-worst-war-crimes-of-the-21st-century/
[5] https://www.kyivpost.com/russias-war/olenivka-prison-camp-massacre-new-details-emerging.html
[6] https://www.bloomberg.com/news/articles/2022-07-20/russians-may-face-first-hague-war-crimes-case-by-end-of-year
[7] https://kse.ua/russia-will-pay/
[8] https://www.worldbank.org/en/country/ukraine/overview

Previous literature has identified ways in which wars can lead to poverty through killing and injuring people, damaging infrastructure, destroying institutions and breaking up communities and networks (Justino 2012). To give one example, Mercier, Ngenzebuke and Verwimp (2020) use a three-wave household-level panel to show that households living in localities exposed to the Burundi Civil War have been subsequently more likely to be poor than non-exposed households. In the context of a transition country, Bisogno and Chong (2002) report a 27.3% poverty rate and an 11.5% rate of extreme poverty in a sample of 20,000 individuals from 66 communities in Bosnia and Herzegovina two years after the war ended.

While the link between unemployment and poverty in developed countries may be mitigated by a generous social security system (Saunders 2002), it is not the case in transition economies. For example, by the end of the war more than half of the Bosnian workforce was unemployed (Giessmann et al 2018). Ten years later the unemployment rate in Bosnia and Herzegovina was 29% and 17.7% of the population was living below the national poverty line.[11] The unemployment rate can also stay high for many years after the conflict, as the example of Kosovo illustrates with its 25.5% unemployment in 2020.[12]

Forced displacement is one of the possible channels for increased unemployment during and after the war. Calderon and Ibanez (2009) indicate that the internal migration caused by the Colombian conflict had a statistically significant positive effect on employment in the informal sector. Kondylis (2010) shows large negative effects of displacement on the employment of Bosnians who left their place of residence after the 1992-1995 war. Torosyan, Pignatti and Obrizan (2018) show that internally displaced people (IDPs) in Georgia could be up to 11.6% points more likely to be unemployed even after two decades of forced displacement.

In addition to individual effects, the destruction of and deferred accumulation of human and physical capital also threatens macroeconomic performance and combines with any effects of war on institutions and technology to impact economic growth (Blattman and Miguel 2010). Rodrik (1999) links the outbreaks of social conflict (which in the case of Ukraine was brought about by an external russian attack) as a primary reason for the lack of persistence in economic growth rates. For example, the war in Sri Lanka had a significant and negative effect estimated at an annual average of 9% of GDP (Ganegodage and Rambaldi 2014). One important clarification should be made – most of the papers considered above focus on civil wars, which is not the case in Ukraine despite the claims of russian propaganda. The conflict in Ukrainian Donbas in 2014 would never have begun had Russia not set foot into Ukrainian territory with its troops.[13]

As such, the paper aims to study the effects of the russian aggression on poverty, unemployment and forced internal displacement in Ukraine three months into the full-scale war. This will add to the literature on the negative socio-economic effects of wars and military conflicts in general and, in particular, on the effects of russian aggression on political participation (Coupe and Obrizan 2016b), happiness (Coupe and Obrizan 2016a) and employment of displaced and non-displaced households (Vakhitova and Iavorskyi 2020) in Donbas as well as job market outcomes of internally displaced people in Georgia (Torosyan, Pignatti and Obrizan 2018), another victim of russian aggression.

The rest of the paper is organized as follows. Section 2 provides information about the survey data and methodology. Section 3 deals with descriptive statistics. Section 4 focuses on the results of regression analyses and robustness checks while the last Section concludes.

**2. Data source and methodology**

---

[9] https://www.undp.org/war-ukraine
[10] https://www.ilo.org/wcmsp5/groups/public/---europe/---ro-geneva/documents/briefingnote/wcms_844295.pdf
[11] https://data.worldbank.org/indicator/SI.POV.NAHC?locations=BA
[12] https://data.worldbank.org/indicator/SL.UEM.TOTL.NE.ZS?locations=XK-BA
[13] https://www.kyivpost.com/article/opinion/op-ed/europeans-subtle-complicity.html

The Omnibus survey data from Kyiv International Institute of Sociology (KIIS) is used in this study including 2,003 and 2,002 observations in two waves of the survey in April 2021 and 2,000 and 2,009 observations in two waves of the survey in May 2022. Hence, there is a healthy number of cross-sectional observations which are collected in the Spring of the two years which should limit the extent of seasonal variation in responses. The Kyiv International Institute of Sociology has a strong reputation as one of the leading survey organizations in Ukraine with its high-quality survey data being used in numerous academic publications (Coupe and Obrizan, 2016a). KIIS Omnibus surveys are conducted via computer-assisted telephone interviews a few times each year from all regions of Ukraine (except the Autonomous Republic of Crimea and non-government-controlled parts of the Donetsk and Luhansk regions) to get a representative sample of the adult Ukrainian population.

Although the effect of the war can be observed all over Ukraine due to launches of kremlin rockets, drones and massive migration, regions directly involved in fighting on the ground are likely to suffer even more. A simple identification strategy is used – regions that have some fighting on the ground are included into the treatment category while the rest of Ukraine is the control group. While having ground fighting in the region is not the same as experiencing it directly, it is still a valid identification strategy given the high population density and well-developed road infrastructure in Ukraine. Once kremlin troops enter the region they can move very quickly as indicated by the evidence in the media of spotting kremlin tanks near Kyiv and other major cities and roads in affected regions from day 1.[14]

This effect is captured by the indicator variable *Ground Attack* which is set to 1 for respondents who lived before February 24th, 2022, in regions that experienced direct fighting on the ground and 0 for other respondents. The regions in the treatment group have been identified using the DeepState Map[15] but are consistent with other sources and include Kyiv, Sumy, Chernihiv, Donetsk, Luhansk, Kharkiv, Zaporizhzhia, Kherson and Mykolaiv regions.[16]

Two socio-economic indicators are considered in this study – self-defined extreme poverty (chosen as the first option in the question on financial status) and unemployment (self-chosen from a list of mutually exclusive occupation options). Other categories for financial status include having enough money for the essentials, having enough money for the essentials but experiencing a problem with the purchase of clothes, three categories for being able to buy expensive things (such as a fridge, a car or housing) and having money to buy everything the household want.[17]

Other categories for occupation include worker (including agricultural workers), clerk (non-physical work not requiring a higher education), specialist (non-physical work requiring a higher education), freelancer, entrepreneur, military serviceman or law enforcement officer, housewife or househusband, retiree, student or other occupations. Hence, the list of occupations includes options for respondents out of the labor force such as students and retirees which will be helpful for the analyses by age.

The analysis begins from a simplistic model which only includes the data for 2022

$$Y_i = \alpha + \beta \cdot X_i + \delta \cdot Ground\ Attack + \varepsilon_i, \quad (1)$$

where $Y_i$ is the dependent variable of interest for respondent $i$ in May of 2022 such as the dummy variable of not having enough money for food or being unemployed. $X_i$ is the vector of socio-demographic and economic characteristics of a respondent $i$ surveyed in May of 2022. While this analysis may be useful to describe the covariates of worsening socio-economic outcomes in 2022 it will not capture the differences between treatment and control groups before and after the full-scale aggression.

---

[14] https://twitter.com/KyivIndependent/status/1497122294258229253
[15] https://deepstatemap.live/#6.5/49.316/31.738
[16] The DeepState Map also identifies very small parts of Zhytomyr and Poltava regions as having direct ground fighting, but they are excluded from the treatment group because of a very small area affected and lack of media coverage of fighting in those two regions compared to others in the treatment group.
[17] The lowest financial category is chosen due to a slight change in the list of options between surveys, but results remain quantitatively similar with a broader definition of poverty although with somewhat reduced statistical significance.

Hence, a difference-in-difference methodology is then applied to study the causal effect of the war similar to Obrizan and Iavorskyi (2022)

$$Y_{i,\ year} = \alpha + \beta \cdot X_{i,\ year} + \gamma \cdot D_{2022} + \delta \cdot Ground\ Attack + \eta \cdot D_{2022} \cdot Ground\ Attack + \varepsilon_{i,year}, \qquad (2)$$

where $D_{2022}$ is an indicator variable taking value of 1 for surveys taken in May of 2022 and 0 for observations in April of 2021 before the full-scale aggression. Observe that $Y_{i,\ year}$ and $X_{i,\ year}$ now include *year* which can be 2021 or 2022. The interaction $D_{2022} \cdot Ground\ Attack$ term will then show the direct effect of the ground conflict in the region on the socio-economic outcome of interest.

## 3. Data description

Table 1 provides descriptive statistics by year and treatment group. The proportion of respondents who did not have enough money for food was 0.076-0.077 in 2021 in both groups. In 2022 the situation somewhat surprisingly improved in the control group with only 0.043 of such respondents but worsened to 0.091 in regions with fighting on the ground. The share of respondents who reported being unemployed was 0.047-0.048 in 2021 in both treatment and control groups. It rose to 0.065 in 2022 in regions without a ground attack and to 0.115 in regions that experienced on-the-ground fighting. Finally, a considerable fraction of 0.266 of respondents in the most affected regions moved after February 23rd, 2022, compared to only 0.032 of respondents in regions further from the frontline.

The descriptive statistics for covariates in Table 1 show a comparable characteristic of respondents except for the share of those who live in mid-sized cities of 100 to 500 thousand people (0.180-0.182 in regions with ground fighting compared to 0.223-0.235 in other regions) and large cities of 500+ thousand respondents (0.363-0.390 in regions with ground fighting compared to 0.206-0.235 in other regions). The higher proportion of respondents in larger cities also manifests itself in a higher share of respondents with higher education (0.440-0.447 in the frontline regions compared to 0.374-0.394 in other regions). Finally, there are no respondents in the Eastern macroregion who are not directly affected because all the regions in the Eastern macroregion experienced fighting on the ground. Likewise, because of the large migration the share of respondents in the Central macroregion is higher in the treatment group and is equal to 0.458-0.503 compared to 0.310-0.315 in the control group not directly affected by the fighting on the ground. Since these issues may potentially affect the stability of the results, they will be addressed in robustness checks.

Table 1. Descriptive statistics by year and treatment group

|  | No ground attack | | Ground attack | |
|---|---|---|---|---|
|  | 2021 | 2022 | 2021 | 2022 |
| Not enough money for food | 0.076 | 0.043 | 0.077 | 0.091 |
|  | (0.264) | (0.204) | (0.267) | (0.287) |
| Unemployed | 0.047 | 0.065 | 0.048 | 0.115 |
|  | (0.212) | (0.247) | (0.215) | (0.319) |
| Moved after Feb 23rd | - | 0.032 | - | 0.266 |
|  | - | (0.177) | - | (0.442) |
| Female | 0.532 | 0.513 | 0.533 | 0.540 |
|  | (0.499) | (0.500) | (0.499) | (0.499) |
| Age | 48.542 | 48.159 | 49.200 | 47.312 |
|  | (16.769) | (16.275) | (16.712) | (16.278) |
| Urban-type settlement | 0.082 | 0.078 | 0.095 | 0.073 |
|  | (0.274) | (0.268) | (0.294) | (0.261) |
| City pop < 20 thousand | 0.075 | 0.081 | 0.056 | 0.058 |
|  | (0.264) | (0.274) | (0.230) | (0.233) |
| City pop > 20 but less than 49 thousands | 0.086 | 0.077 | 0.086 | 0.080 |
|  | (0.280) | (0.267) | (0.280) | (0.271) |
| City pop > 50 but less than 100 thousands | 0.070 | 0.054 | 0.078 | 0.061 |
|  | (0.256) | (0.227) | (0.268) | (0.239) |
| City pop > 100 but less | 0.223 | 0.235 | 0.180 | 0.182 |
| than 500 thousands                 | (0.416) | (0.424) | (0.384) | (0.386) |
| City pop > 500 thousands           | 0.206   | 0.235   | 0.363   | 0.390   |
|                                    | (0.404) | (0.424) | (0.481) | (0.488) |
| High school dropout                | 0.029   | 0.016   | 0.024   | 0.019   |
|                                    | (0.167) | (0.124) | (0.154) | (0.138) |
| Vocational without high school diploma | 0.025 | 0.027 | 0.020 | 0.024 |
|                                    | (0.155) | (0.161) | (0.140) | (0.152) |
| High school                        | 0.142   | 0.138   | 0.116   | 0.114   |
|                                    | (0.349) | (0.345) | (0.320) | (0.317) |
| Vocational with high school diploma | 0.053  | 0.053   | 0.045   | 0.042   |
|                                    | (0.225) | (0.225) | (0.208) | (0.200) |
| Specialized vocational             | 0.311   | 0.293   | 0.274   | 0.279   |
|                                    | (0.463) | (0.455) | (0.446) | (0.449) |
| Incomplete higher education        | 0.063   | 0.069   | 0.066   | 0.074   |
|                                    | (0.243) | (0.254) | (0.249) | (0.262) |
| Higher education                   | 0.374   | 0.394   | 0.447   | 0.440   |
|                                    | (0.484) | (0.489) | (0.497) | (0.496) |
| Central macroregion                | 0.310   | 0.315   | 0.458   | 0.503   |
|                                    | (0.462) | (0.465) | (0.498) | (0.500) |
| Southern macroregion               | 0.299   | 0.299   | 0.234   | 0.217   |
|                                    | (0.458) | (0.458) | (0.423) | (0.413) |
| Eastern macroregion                | 0.000   | 0.000   | 0.308   | 0.280   |
|                                    | (0.000) | (0.000) | (0.462) | (0.449) |
| Number of observations             | 2103    | 2099    | 1902    | 1910    |

*Notes: Author's calculations based on KIIS Omnibus surveys. The base settlement category is living in a village with an urban-type settlement being a category in between a village and a city. The base category for education is basic (7 grades). The base category for macroregion is the Western macroregion.*

A significant deterioration of socio-economic outcomes in regions with direct ground fighting may be partially explained by the changes in observable characteristics of the respondents. It is time to proceed to regression analyses that will take into account those changes.

**4. Results**

The first regression model (1) includes only 2022 in order to identify covariates of the worsening socio-economic situation. The results reported in Table 2 show that respondents in the treatment group had 3.6-3.8% points higher chances of not having enough money for food and 2.0-3.1% points higher probability of being unemployed with the coefficient for *Ground Attack* being significant at 1%. Variable *Moving after February 23rd* did not have a separate statistically significant effect on self-reported extreme poverty but is associated with additional 7.5% points higher probability of self-reported unemployment. Finally, the respondents in the treatment group are 13.9% points more likely to become displaced after February 23rd, 2022.

Other coefficients significant at 1% include Females who are more likely to report extreme poverty by 2.4% points. Older respondents are more likely to not have enough money for food and less likely to be unemployed (there is a separate category for retirees) or to relocate after February 23rd, 2022. While all of the coefficients for Age are statistically significant at 1%, they have a very small effect in the range of 0.001-0.002 for one additional year of life. Respondents in the Southern and Eastern macroregions are 4.0-6.7% points more likely to report extreme poverty compared to the Western macroregion. Respondents in the Southern and Eastern macroregions are also more likely to become unemployed by 2.2-2.5% points (significant at 5%) and 5.5-8.1% points correspondingly.

Table 2. Determinants of socio-economic outcomes in 2022

| | No Money | No Money | Unemployed | Unemployed | Moved After |
|---|---|---|---|---|---|

|  | for Food | for Food |  |  | Feb 23rd |
|---|---|---|---|---|---|
| Ground Attack | 0.036*** | 0.038*** | 0.020*** | 0.031*** | 0.139*** |
|  | (0.008) | (0.008) | (0.007) | (0.007) | (0.026) |
| Moved after Feb 23rd |  | 0.009 |  | 0.075*** |  |
|  |  | (0.008) |  | (0.018) |  |
| Female | 0.024*** | 0.024*** | -0.011 | -0.010 | 0.015 |
|  | (0.006) | (0.006) | (0.009) | (0.009) | (0.009) |
| Age | 0.001*** | 0.001*** | -0.002*** | -0.002*** | -0.002*** |
|  | (0.000) | (0.000) | (0.000) | (0.000) | (0.000) |
| City pop < 20 thousand | -0.030* | -0.029* | 0.008 | 0.010 | 0.018 |
|  | (0.015) | (0.014) | (0.018) | (0.019) | (0.027) |
| City pop > 100 but less than 500 thousands | -0.013 | -0.012 | -0.010 | -0.006 | 0.063** |
|  | (0.020) | (0.019) | (0.010) | (0.010) | (0.024) |
| City pop > 500 thousands | -0.034** | -0.034** | -0.006 | -0.006 | 0.002 |
|  | (0.015) | (0.015) | (0.010) | (0.010) | (0.027) |
| High school dropout | 0.120* | 0.120* | 0.005 | 0.004 | -0.013 |
|  | (0.065) | (0.064) | (0.042) | (0.045) | (0.060) |
| High school | 0.067 | 0.066 | 0.000 | -0.006 | -0.079* |
|  | (0.040) | (0.040) | (0.037) | (0.039) | (0.043) |
| Southern microregion | 0.040*** | 0.040*** | 0.022** | 0.025** | 0.039 |
|  | (0.007) | (0.007) | (0.010) | (0.010) | (0.024) |
| Eastern microregion | 0.064*** | 0.067*** | 0.055** | 0.081*** | 0.355*** |
|  | (0.015) | (0.013) | (0.023) | (0.019) | (0.093) |
| Constant | -0.035 | -0.034 | 0.194*** | 0.205*** | 0.148*** |
|  | (0.046) | (0.046) | (0.035) | (0.038) | (0.042) |
| Observations | 4009 | 4009 | 4009 | 4009 | 4009 |
| Adjusted R-squared | 0.042 | 0.042 | 0.043 | 0.036 | 0.229 |

*Notes: * p<0.10, ** p<0.05, *** p<0.01. Standard errors in parentheses. Estimated by OLS with robust standard errors clustered at the level of region (oblast). All models control for the list of covariates specified in Table 1 but only statistically significant coefficients are shown.*

Although these descriptive regression results are informative, they do not have a causal interpretation because they do not reflect the potential differences between the treatment and control groups that existed before 2022. The descriptive statistics in Table 1 show some differences in observable characteristics related to the city of residence and education. Hence, the analysis proceeds to the preferred difference-in-difference specification in Table 3.

The results in Table 3 indicate that regions that experienced direct kremlin attacks on the ground were not different before the full-scale aggression from other regions in terms of extreme poverty and unemployment. Extreme poverty somewhat reduced by 2.8-3.2% points in 2022 in regions not directly affected by fighting - an observation that was already seen in descriptive statistics. Most importantly, the probability to lack money for food increased because of the full-scale war by 4.6% points and this result is remarkably stable in a model with only difference-in-difference indicators and the full set of covariates. Being a Female is associated with an evaluated risk of not having enough money for food by 3.5% points and older respondents are more likely to experience poverty, but the quantitative effect is again small. The effect of other covariates remains mostly insignificant or marginally significant at a 5% or 10% significance level except for reduced risks of poverty in cities with less than 20 thousand people and elevated chances of poverty in the Southern and Eastern macroregions by 3.1-4.1% points.

Table 3. The results of the difference-in-difference estimation

|  | No Money for Food | No Money for Food | Unemployed | Unemployed |
|---|---|---|---|---|
| Ground Attack | 0.002 | -0.003 | 0.001 | -0.006 |
|  | (0.010) | (0.010) | (0.008) | (0.010) |
| Year 2022 | -0.032*** | -0.028*** | 0.018* | 0.019** |

|  | (0.005) | (0.006) | (0.009) | (0.009) |
|---|---|---|---|---|
| Year 2022*Ground Attack | 0.046*** | 0.046*** | 0.048*** | 0.045*** |
|  | (0.010) | (0.011) | (0.014) | (0.014) |
| Female |  | 0.035*** |  | -0.001 |
|  |  | (0.007) |  | (0.005) |
| Age |  | 0.002*** |  | -0.002*** |
|  |  | (0.000) |  | (0.000) |
| City pop < 20 thousand |  | -0.029*** |  | -0.007 |
|  |  | (0.010) |  | (0.013) |
| City pop > 20 but less than 49 thousands |  | -0.029** |  | -0.012 |
|  |  | (0.012) |  | (0.014) |
| City pop > 50 but less than 100 thousands |  | -0.007 |  | -0.022* |
|  |  | (0.014) |  | (0.013) |
| City pop > 500 thousands |  | -0.023* |  | -0.018* |
|  |  | (0.011) |  | (0.009) |
| High school dropout |  | 0.098* |  | 0.030 |
|  |  | (0.050) |  | (0.027) |
| Southern macroregion |  | 0.031*** |  | 0.018** |
|  |  | (0.008) |  | (0.008) |
| Eastern macroregion |  | 0.041*** |  | 0.047*** |
|  |  | (0.009) |  | (0.014) |
| Constant | 0.076*** | -0.053 | 0.047*** | 0.148*** |
|  | (0.006) | (0.041) | (0.006) | (0.022) |
| Observations | 8014 | 8014 | 8014 | 8014 |
| Adjusted R-squared | 0.004 | 0.043 | 0.011 | 0.032 |

*Notes: * p<0.10, ** p<0.05, *** p<0.01. Standard errors in parentheses. Estimated by OLS with robust standard errors clustered at the level of region (oblast). All models control for the list of covariates specified in Table 1 but only statistically significant coefficients are shown.*

Additional models in Table 4 explore heterogeneity by gender and the presence of higher education. The results imply that extreme poverty due to the full-scale war is only observed in the sample of females without higher education with risks increased by 9.6-9.9% points. The effect remains insignificant in the other three groups – females with higher education and both groups of men with and without higher education. This provides clear policy guidance in terms of providing basic food assistance to females without higher education.

Table 4. Exploring heterogeneity by gender and education

|  | No Money for Food | No Money for Food | Unemployed | Unemployed |
|---|---|---|---|---|
|  | *Females without higher education* | | | |
| Ground Attack | -0.018 | -0.037 | -0.002 | -0.007 |
|  | (0.022) | (0.026) | (0.009) | (0.011) |
| Year 2022 | -0.059*** | -0.053*** | 0.003 | 0.005 |
|  | (0.015) | (0.014) | (0.016) | (0.015) |
| Year 2022*Ground Attack | 0.096*** | 0.099*** | 0.069*** | 0.061*** |
|  | (0.022) | (0.024) | (0.021) | (0.019) |
| Full set of covariates | No | Yes | No | Yes |
| Observations | 2472 | 2472 | 2472 | 2472 |
| Adjusted R-squared | 0.007 | 0.034 | 0.011 | 0.037 |
|  | No Money for Food | No Money for Food | Unemployed | Unemployed |
|  | *Females with higher education* | | | |
| Ground Attack | 0.005 | -0.006 | -0.006 | -0.018 |
|  | (0.016) | (0.020) | (0.016) | (0.016) |
| Year 2022 | -0.042** | -0.044** | 0.003 | 0.003 |
|  | (0.016) | (0.016) | (0.017) | (0.016) |
| Year 2022*Ground Attack | 0.024 | 0.026 | 0.047** | 0.042** |
|  | (0.025) | (0.025) | (0.021) | (0.020) |
| Full set of covariates | No | Yes | No | Yes |
| Observations | 1769 | 1769 | 1769 | 1769 |
| Adjusted R-squared | 0.006 | 0.027 | 0.006 | 0.031 |
|  | No Money for Food | No Money for Food | Unemployed | Unemployed |
|  | *Males without higher education* | | | |
| Ground Attack | 0.035*** | 0.033*** | -0.008 | -0.009 |
|  | (0.010) | (0.012) | (0.017) | (0.019) |
| Year 2022 | -0.006 | -0.005 | 0.043*** | 0.043*** |
|  | (0.011) | (0.012) | (0.012) | (0.011) |
| Year 2022*Ground Attack | 0.017 | 0.017 | 0.065*** | 0.066*** |
|  | (0.017) | (0.017) | (0.021) | (0.021) |
| Full set of covariates | No | Yes | No | Yes |
| Observations | 2237 | 2237 | 2237 | 2237 |
| Adjusted R-squared | 0.007 | 0.017 | 0.020 | 0.029 |
|  | No Money for Food | No Money for Food | Unemployed | Unemployed |
|  | *Males with higher education* | | | |
| Ground Attack | 0.004 | 0.007 | 0.035*** | 0.007 |
|  | (0.010) | (0.014) | (0.010) | (0.017) |
| Year 2022 | -0.004 | -0.002 | 0.024* | 0.023 |
|  | (0.009) | (0.009) | (0.014) | (0.015) |
| Year 2022*Ground Attack | 0.021 | 0.024 | -0.000 | -0.003 |
|  | (0.018) | (0.019) | (0.024) | (0.023) |
| Full set of covariates | No | Yes | No | Yes |
| Observations | 1536 | 1536 | 1536 | 1536 |
| Adjusted R-squared | 0.001 | 0.007 | 0.007 | 0.023 |

*Notes: * p<0.10, ** p<0.05, *** p<0.01. Standard errors in parentheses. Estimated by OLS with robust standard errors clustered at the level of region (oblast). All models with a full set of covariates control for the list of variables specified in Table 1.*

The situation with unemployment is somewhat different. Table 3 shows that the respondents in regions with fighting on the ground have reported a probability of being unemployed higher by 4.5-4.8% points depending on the model specification. Here there is no statistically significant effect for being a female and, while older respondents are less likely to be unemployed, the quantitative effect is close to zero. The only other coefficient which is significant at 1% is living in the Eastern macroregion before the full-scale far which resulted in 4.7% points higher probability of self-reported unemployment.

When the focus turns to Table 4 one can observe that only men with higher education did not experience an elevated chance to become unemployed due to the war. Females with higher education report a smaller magnitude of unemployment probability by 4.2-4.7% points due to the full-scale war (significant at 5%) but it is comparable in magnitude to 6.1-6.9% points for females without higher education. The gender gap manifests itself in the fact that higher education does not fully protect females against unemployment! In particular, males without higher education face a similar increase in unemployment (6.5-6.6% points) comparable to females with higher education.

Table A1 in the Appendix in addition shows that the causal effect of the full-scale aggression on regions with a ground attack remains significant at 1% in the model including only females (5.4-5.9% points) and is marginally significant at 10% in the sample including only males (3.5-3.7% points). Here higher education is protective against unemployment with an insignificant effect of the war, but respondents without higher education report a 6.5-6.6% points higher probability of being unemployed due to the full-scale aggression.

A number of additional robustness checks are considered in Table A1 in Appendix. The negative effects of the full-scale war on poverty are only observed in the sample with females (by 6.6-6.7% points) but not males and only for respondents without higher education (by 6.0-6.3% points) but not for respondents with higher education. The model without macroregions shows a 4.5-4.6% points higher probability of not having enough money for food and 4.4-4.8% points higher risks of being unemployed for respondents in directly affected regions with ground fighting. Negative effects of the war also remain robust in samples divided by a median sample age of 48, although with a lower significance in the older subsample. Hence, the results remain robust in these additional models.

## 5. Discussion and Conclusions

This paper adds to the existing literature on the effect of military conflict on population poverty and unemployment and is one of the first to establish the causal effect of the full-scale russian aggression on the socio-economic status of Ukrainians who experienced brutal fighting on the ground in their home region. The descriptive regressions show that moving after February 23rd to another location does not have a separate effect on extreme poverty but increases the probability of becoming unemployed by 7.5% points in 2002. The causal models, on the other hand, indicate a strong gender gap in the effect of the full-scale war. In particular, women without higher education are the only group with a statistically significant higher probability of extreme poverty. Thus, any programs of social assistance to the poor during the war should focus on this group in the first place.

In terms of unemployment, the only unaffected group is men with a higher education while being a female with a higher education still does not fully protect from the risk of unemployment and has a comparable magnitude of risk to that of men without a higher education. This contradicts recent evidence on the lack of "she-cession" during COVID-19 in Ukraine in Brik and Obrizan (2022) which was linked to telecommuting as a way to mitigate the unemployment risk. By May 2022 many Ukrainian women with children remained internally displaced which could have limited their ability to preserve jobs over long distance even with the possibility of telecommuting.

The presented paper clearly demonstrates the causal effect of the war on worsening socio-economic outcomes of Ukrainian citizens, but it is not without limitations that warrant further discussion. First, the survey in May of 2022 excludes respondents who left Ukraine after February 23rd, 2022. However, the descriptive statistics in Table 1 do not indicate a substantial difference between the respondents in 2021 and 2022 in terms of their observable characteristics. In addition, the Omnibus survey does not include many characteristics that may affect the probability of being poor or unemployed. But the results remain remarkably stable with only the set of difference-in-difference indicators, in a model with the full list of covariates and also in the robustness checks. Finally, kremlin rocket and drone terror of the regions not on the frontline has intensified during the summer of 2022 which cannot be reflected in the data from May 2022. While more accurate results may obtain once more comprehensive surveys become available, this paper provides a stable initial estimate of the war effects on poverty, unemployment and internal migration three months into the full-scale military conflict.

## References

Bisogno, M. and Chong, A. 2002. Poverty and inequality in Bosnia and Herzegovina after the civil war. *World development*, 30(1), pp.61-75.

Blattman, C. and Miguel, E. 2010. Civil war. *Journal of Economic literature*, 48(1), pp. 3-57.

Brik, T. and Obrizan, M. 2022. Gender gap in urban job market during the pandemic – the case of Ukraine. *Revise & Resubmit to Comparative Economic Studies*.

Calderon, V. and Ibanez, A. 2009. Labor market effects of migration-related supply shocks: evidence from internally displaced populations in Colombia. MICROCON Research Working Paper 14.

Coupe, T. and Obrizan, M. 2016a. The impact of war on happiness: The case of Ukraine. *Journal of Economic Behavior & Organization*, 132, pp. 228-242.

Coupe, T. and Obrizan, M. 2016b. Violence and political outcomes in Ukraine—evidence from Sloviansk and Kramatorsk. *Journal of Comparative Economics*, 44(1), pp.201-212.

Ganegodage, K.R. and Rambaldi, A.N. 2014. Economic consequences of war: Evidence from Sri Lanka. *Journal of Asian Economics*, 30, pp.42-53.

Giessmann, H.J., Mac Ginty, R., Austin, B. and Seifert, C. eds., 2018. *The Elgar Companion to Post-conflict Transition*. Edward Elgar Publishing.

Justino, P. 2012. War and poverty. *IDS Working Papers*, 2012(391), pp. 1-29.

Kondylis, F. 2010. Conflict displacement and labor market outcomes in post-war Bosnia and Herzegovina. *Journal of Development Economics*, 93, pp. 235–248.

Mercier, M., Ngenzebuke, R.L. and Verwimp, P., 2020. Violence exposure and poverty: Evidence from the Burundi civil war. *Journal of Comparative Economics*, 48(4), pp.822-840.

Obrizan, M. and Iavorskyi, P. 2022. Health Consequences of the War in Eastern Ukraine: Comparing 2015-16 to 2012-13. *Revise & Resubmit in European Journal of Comparative Economics*.

Rodrik, D. 1999. Where did all the growth go? External shocks, social conflict, and growth collapses. *Journal of economic growth*, 4(4), pp.385-412.

Saunders, P., 2002. The direct and indirect effects of unemployment on poverty and inequality. *Australian Journal of Labour Economics*, 5(4), pp.507-529.

Torosyan, K., Pignatti, N. and Obrizan, M. 2018. Job market outcomes for IDPs: The case of Georgia. *Journal of Comparative Economics*, 46(3), pp.800-820.

Vakhitova, H. and Iavorskyi, P. 2020. Employment of Displaced and Non-Displaced Households in Luhansk and Donetsk Oblasti. *Europe-Asia Studies*, 72(3), pp.383-403.

World Bank. 2022. "War in the Region" Europe and Central Asia Economic Update (Spring), Washington, DC: World Bank. Doi: 10.1596/978-1-4648-1866-0.

**Online Appendix**

Table A1. Robustness checks

|  | No Money for Food | No Money for Food | Unemployed | Unemployed |
|---|---|---|---|---|
|  | *Females only* | | | |
| Ground Attack | -0.014 | -0.024 | -0.005 | -0.013 |
|  | (0.019) | (0.021) | (0.008) | (0.009) |
| Year 2022 | -0.055*** | -0.049*** | 0.002 | 0.004 |
|  | (0.011) | (0.011) | (0.011) | (0.010) |
| Year 2022*Ground Attack | 0.067*** | 0.066*** | 0.059*** | 0.054*** |
|  | (0.017) | (0.017) | (0.015) | (0.014) |
| Full set of covariates | No | Yes | No | Yes |
| Observations | 4241 | 4241 | 4241 | 4241 |
| Adjusted R-squared | 0.005 | 0.049 | 0.009 | 0.036 |
|  | No Money for Food | No Money for Food | Unemployed | Unemployed |
|  | *Males only* | | | |
| Ground Attack | 0.020** | 0.023** | 0.009 | 0.001 |
|  | (0.007) | (0.009) | (0.011) | (0.015) |
| Year 2022 | -0.005 | -0.004 | 0.036*** | 0.035*** |
|  | (0.007) | (0.007) | (0.010) | (0.009) |
| Year 2022*Ground Attack | 0.020 | 0.021 | 0.037* | 0.035* |
|  | (0.014) | (0.014) | (0.019) | (0.019) |
| Full set of covariates | No | Yes | No | Yes |
| Observations | 3773 | 3773 | 3773 | 3773 |
| Adjusted R-squared | 0.005 | 0.023 | 0.013 | 0.029 |
|  | No Money for Food | No Money for Food | Unemployed | Unemployed |
|  | *With higher education* | | | |
| Ground Attack | 0.005 | -0.000 | 0.013 | -0.006 |
|  | (0.008) | (0.011) | (0.012) | (0.014) |
| Year 2022 | -0.024*** | -0.024*** | 0.013 | 0.012 |
|  | (0.006) | (0.007) | (0.013) | (0.014) |
| Year 2022*Ground Attack | 0.022 | 0.024 | 0.025 | 0.021 |
|  | (0.016) | (0.016) | (0.020) | (0.019) |
| Full set of covariates | No | Yes | No | Yes |

|  | No Money for Food | No Money for Food | Unemployed | Unemployed |
|---|---|---|---|---|
| Observations | 3305 | 3305 | 3305 | 3305 |
| Adjusted R-squared | 0.003 | 0.021 | 0.006 | 0.028 |
| | *Without higher education* | | | |
| Ground Attack | 0.006 | -0.005 | -0.005 | -0.008 |
|  | (0.011) | (0.013) | (0.009) | (0.010) |
| Year 2022 | -0.036*** | -0.032*** | 0.022* | 0.023** |
|  | (0.007) | (0.007) | (0.012) | (0.011) |
| Year 2022*Ground Attack | 0.060*** | 0.063*** | 0.066*** | 0.065*** |
|  | (0.011) | (0.012) | (0.017) | (0.016) |
| Full set of covariates | No | Yes | No | Yes |
| Observations | 4709 | 4709 | 4709 | 4709 |
| Adjusted R-squared | 0.006 | 0.037 | 0.015 | 0.033 |
|  | No Money for Food | No Money for Food | Unemployed | Unemployed |
| | *Exclude macroregions* | | | |
| Ground Attack | 0.002 | 0.007 | 0.001 | 0.007 |
|  | (0.010) | (0.009) | (0.008) | (0.007) |
| Year 2022 | -0.032*** | -0.028*** | 0.018* | 0.019** |
|  | (0.005) | (0.006) | (0.009) | (0.009) |
| Year 2022*Ground Attack | 0.046*** | 0.045*** | 0.048*** | 0.044*** |
|  | (0.010) | (0.011) | (0.014) | (0.014) |
| Full set of covariates | No | Yes | No | Yes |
| Observations | 8014 | 8014 | 8014 | 8014 |
| Adjusted R-squared | 0.004 | 0.041 | 0.011 | 0.030 |
|  | No Money for Food | No Money for Food | Unemployed | Unemployed |
| | *Age below median* | | | |
| Ground Attack | -0.003 | -0.007 | 0.008 | -0.000 |
|  | (0.009) | (0.011) | (0.013) | (0.015) |
| Year 2022 | -0.012 | -0.011 | 0.019 | 0.021 |
|  | (0.008) | (0.008) | (0.015) | (0.015) |
| Year 2022*Ground Attack | 0.056*** | 0.053*** | 0.065*** | 0.062*** |
|  | (0.014) | (0.015) | (0.021) | (0.021) |
| Full set of covariates | No | Yes | No | Yes |
| Observations | 3897 | 3897 | 3897 | 3897 |
| Adjusted R-squared | 0.010 | 0.037 | 0.015 | 0.026 |
|  | No Money for Food | No Money for Food | Unemployed | Unemployed |
| | *Age above median* | | | |
| Ground Attack | 0.005 | -0.001 | -0.004 | -0.010 |
|  | (0.014) | (0.014) | (0.008) | (0.009) |
| Year 2022 | -0.050*** | -0.044*** | 0.017** | 0.015*** |
|  | (0.008) | (0.009) | (0.007) | (0.005) |
| Year 2022*Ground Attack | 0.040** | 0.039** | 0.026* | 0.026** |
|  | (0.017) | (0.018) | (0.013) | (0.012) |
| Full set of covariates | No | Yes | No | Yes |
| Observations | 4117 | 4117 | 4117 | 4117 |
| Adjusted R-squared | 0.005 | 0.032 | 0.005 | 0.045 |

*Notes: * p<0.10, ** p<0.05, *** p<0.01. Standard errors in parentheses. Estimated by OLS with robust standard errors clustered at the level of region (oblast). All models with a full set of covariates control for the list of variables specified in Table 1.*